**Highly Accurate Solutions of the Blasius and Falkner-Skan Boundary Layer Equations via Convergence Acceleration**

B.D. Ganapol
Department of Aerospace and Mechanical Engineering
University of Arizona

## ABSTRACT
A new highly accurate algorithm for the solution of the Falkner-Skan equation of boundary layer theory is presented. The algorithm, based on a Maclaurin series representation, finds its coefficients from recurrence. In addition, Wynn-epsilon convergence acceleration and continuous analytical continuation enable an accurate evaluation. The most accurate skin friction coefficients (shooting angle) to date are presented along with comparisons to past and present values found in the literature. The algorithm, coded in FORTRAN, uses neither enhanced precision arithmetic beyond quadruple precision nor computer algebra to achieve results in a timely fashion.

## 1. INTRODUCTION
Every since its first appearance in the literature in 1908 [1], the Blasius equation describing viscous flow over a flat plate has fascinated physicists, engineers, mathematicians and numerical analysts alike. This ODE is rich in physical, mathematical and numerical challenges. Two-dimensional flow over a fixed impenetrable surface creates a boundary layer as particles move more slowly near the surface than near the free stream. Because of its application to fluid flow, physicists and engineers have a keen interest in solving the Blasius equation and the related, but more general, Falkner-Skan (F-S) equation [2]. Since one can elegantly reduce these equations to one-dimensional non-linear ODEs through similarity arguments, mathematicians have found their fulfillment in uncovering the underlying symmetries and proving existence and (non-) uniqueness of its solutions. Unfortunately, a general analytical solution has not been forthcoming; however, for special cases of the F-S equation, several analytical solutions do exist. These prove most beneficial in verifying numerical algorithms. Numerical analysts, or as they are called by J.P Boyd, “arithmurgists” [3], have had a “field-day with these equations. They offer the mystery of nonlinearity— yet the simplicity of one-dimension—and the challenge of solving a boundary value problem through the determinism of an initial value problem. A host of numerical methods have emerged to solve these equations including, but not limited to, finite differences and finite elements, Adomian's polynomials, perturbation methods, homotopy, differential transformation, generalized Laguerre and Chebyshev polynomial expansions, variational iteration, the quasi-linear approximation and the diagonal Padé approximant. A list of nearly 150 references (some repeated) for these and other algorithms is found in Refs. 4 and 5. We also note that the majority of the numerical algorithms are based on Runge-Kutta ODE solvers, and therefore have a definite "black box" quality. In spite of the enormous numerical effort however, a truly simple, yet numerically accurate and robust algorithm is still missing. Many, if not all, algorithms to this point seem rather delicate in that their iterative strategies must be carefully tuned to avoid numerical instability. For example, most schemes require the initial guess of the shooting angle to be relatively close to the converged result, which does not make for a robust algorithm. Judging from recent literature, the general lack of numerical agreement to a consistent five or more digits is indicative of the need for a reliable algorithm--the development of which we now address.

The goal of the present investigation is to provide a robust algorithm for the solution of the F-S equation for physically relevant flows and for some additional flows found in the literature. We report highly accurate solutions to at least 10 or more digits without resorting to enhanced precision arithmetic or computer algebra. All calculations are programmed in FORTRAN double precision (DP) arithmetic with the exception of several quadruple precision (QP) demonstrations of claimed accuracy.

We consider the F-S equation,

$$f'''(\eta)+\beta_0 f(\eta)f''(\eta)+\beta\left(1-f'(\eta)^2\right)=0,\quad \eta\in[0,\infty), \tag{1a}$$

Physically, the F-S equation describes 2D flow over stationary impenetrable wedge surfaces of included angle $\beta\pi$, which limits to a flat plate, and the Blasius solution, as $\beta$ approaches zero. Eq(1a) is subject to the following boundary conditions:

$$\begin{aligned} &f(0)=0\\ &f'(0)=0\\ &\lim_{\eta\to\infty} f'(\eta)=1, \end{aligned} \tag{1b}$$

where the velocity profile goes asymptotically to unity. While $\beta_{\min}$ □ $37735 < \beta < 2$ represents physical flow, with two solutions in the negative range of $\beta$ for forward and reverse flow, solutions for larger values of $\beta$, as well, found in the literature, are also considered.

Curiously, as will be shown, we find that the most numerically accurate form of solution was originally proposed by Blasius himself in 1908 [1] as the Maclaurin series

$$f(\eta) = \sum_{k=0}^{\infty} \frac{f^{(k)}(0)}{k!} \eta^k . \tag{2}$$

Moreover, all the elements of a particularly robust solution, exhibiting superior accuracy already exist in the literature. The relations between the elements along with some numerical common sense then provide the desired robust numerical algorithm we seek.

In the following section, we derive the fundamental algorithm based on recurrence for the derivatives in the solution representation of Eq(2). Note that the recurrence was derived in Ref. 6 in an equivalent vector form, but the chosen implementation limited the method's full potential. Even with its coefficients known, direct evaluation of Eq(2) is problematic because of the finite radius of convergence of the series representation [7]. For this reason, as purposed in Ref. 8, we apply a Pade approximant. Unlike that suggested however, we seamlessly implement the diagonal approximant through Wynn-epsilon convergence acceleration. Unfortunately, convergence acceleration alone is insufficient to give the desired high accuracy. To achieve high accuracy therefore, continuous analytical continuation (CAC), applied in the manner suggested also in Ref. 6, is the remedy. Then, with confidence in our numerical evaluation, we are in position to implement the shooting method taking advantage of the physical features of the velocity profile as identified in Ref 9. We conclude with comparisons to literature values and, when convenient, add digits to establish new benchmark standards.

Therefore as indicated, an accurate algorithm for resolution of the F-S equation has existed in the literature since the application of Wynn-epsilon convergence acceleration [10]. Its emergence however, required reliable computing, courage to embrace experimental mathematics and an extreme numerical desire.

## 2. METHOD OF SOLUTION

In this section, we derive the recurrence for the derivatives of Eq(2) and formulate the shooting method.

### 2.1 Maclaurin series representation

A recurrence relation for the derivatives in the Maclaurin series is found by applying the differential operator [6]

$$\frac{d^k}{d\eta^k}$$

to Eq(1a). In the process, note that

$$\frac{d^k}{d\eta^k} f(\eta) f''(\eta) = \sum_{l=0}^{k} \frac{k!}{l!(k-l)!} f^{(l+2)}(\eta) f^{(k-l)}(\eta)$$

and

$$\frac{d^k}{d\eta^k} f(\eta)^2 = \sum_{l=0}^{k} \frac{k!}{l!(k-l)!} f^{(l+1)}(\eta) f^{(k-l+1)}(\eta)$$

to give

$$\begin{aligned} f^{(k+3)}(\eta) + \beta_0 \sum_{l=0}^{k} \frac{k!}{l!(k-l)!} f^{(l+2)}(\eta) f^{(k-l)}(\eta) - \\ -\beta \sum_{l=0}^{k} \frac{k!}{l!(k-l)!} f^{(l+1)}(\eta) f^{(k-l+1)}(\eta) + \beta \delta_{k0} = 0. \end{aligned} \tag{3}$$

If

$$f^{(k)}(\eta) = k! a_k ,$$

then, with some algebra

$$a_k(\eta) = \frac{1}{k(k-1)}\left[\beta a_1(\eta)a_{k-2}(\eta) - \beta_0(k-1)a_0(\eta)a_{k-1}(\eta)\right] - \frac{\beta}{6}\delta_{k,3} + $$
$$+\frac{1}{k(k-1)(k-2)}\sum_{l=2}^{k-2} l\left[\beta(k-l-1) - \beta_0(l-1)\right]a_l(\eta)a_{k-l-1}(\eta), \tag{4a}$$

and the boundary conditions become

$$a_0(0) = 0$$
$$a_1(0) = 0$$
$$\lim_{\eta\to\infty} a_1(\eta) = 1.$$

Thus,

$$f(\eta) = \sum_{k=0}^{\infty} a_k(0)\eta^k \tag{5}$$

is the desired evaluation.

There are three issues to emphasize here. First, the dependent variables remain in their original domain and are not in a finite interval as found in many of the previous investigations. Such a transformation gives a false sense of security of improved accuracy because one has avoided the infinity of $\eta$. With a transformation however comes the need to include an additional boundary condition, which unnecessarily complicates the shooting strategy by requiring an optimum search in a 2D parameter space (for the shooting angle and the additional condition). The second point is that by formulating the solution in terms of derivatives only, the ODE's nonlinearity is accommodated naturally through the (nonlinear) recurrence itself. There is no need to explicitly consider the nonlinearity of Eq(1a) any further. Finally, we are not reliant on an ODE solver with its inherent truncation and roundoff errors. While marching is still required, it is more accurately accomplished by evaluating power series solutions over contiguous intervals.

## 2.2 THE SHOOTING METHOD

The shooting method is the preferred way to treat the F-S boundary value problem. Through a specification of an additional initial condition to replace the condition at infinity, the boundary value problem transforms into an equivalent iterative initial value problem. The initial condition is $f''(0,\alpha)=\alpha$, where $\alpha$ is the skin friction coefficient. To be equivalent, the shooting angle $\alpha$ must be determined such that $\lim_{\eta\to\infty} f'(\eta;\alpha)=1$. Except where noted for a particular range of $\beta$, the solution is assumed unique [12]. Now, the recurrence begins with

$$\begin{aligned} a_0(0,\alpha)&=0\\ a_1(0,\alpha)&=0\\ a_2(0,\alpha)&=\alpha/2. \end{aligned} \tag{4b}$$

In this way, Eq(5), with $\alpha$ assumed, gives $f$ and its first two derivatives as

$$\begin{aligned} f(\eta;\alpha)&=\sum_{k=0}^{\infty} a_k(0,\alpha)\eta^k\\ f'(\eta;\alpha)&=\sum_{k=0}^{\infty} k a_k(0,\alpha)\eta^{k-1}\\ f''(\eta;\alpha)&=\sum_{k=0}^{\infty} k(k-1)a_k(0,\alpha)\eta^{k-2}, \end{aligned} \tag{7}$$

where a prime refers to differentiation with respect to $\eta$. Note the explicit dependence of $\alpha$ since $\alpha$ is to be varied.

## 3. NUMERICAL IMPLEMENTATION OF THE SERIES SOLUTION REPRESENTATION

We now consider how to best evaluate Eq(5) to a desired accuracy, which will be through convergence acceleration.

### 3.1 Evaluation of the Maclaurin series by convergence acceleration

It is well known that the radius of convergence for the series in Eq(5) is finite, thus limiting its usefulness. For this reason, we apply Wynn-epsilon convergence to accelerate the partial sums to convergence.

First, consider the evaluation of a general infinite series

$$S(\eta)=\sum_{k=0}^{\infty} s_k\eta^k,$$

which is, in the limit

$$S(\eta) = \lim_{l \to \infty} S_l(\eta),$$

if it exists, where the partial sums are

$$S_l(\eta) = \sum_{k=0}^{l} s_k \eta^k .$$

In this way, a sequence of approximations to the limit

$$\mathbf{S} \equiv \left\{ S_l(\eta), l = 1, 2, .... \right\},$$

is created and assumed to uniformly converge.

To accelerate convergence, one applies the Wynn-epsilon (*We*) [11] convergence accelerator, which frequently results in a more rapidly converging sequence than the original. In this regard, we enter into the realm of experimental numerical methods since there is no proof that the F-S series solution lends itself to such an acceleration. There is some indication of success however, in that a diagonal Pade approximant gives a solution [8] and the *We* algorithm is just that [13].

The *We* accelerator takes the following form for the sequence $S_l(\eta);\ l = 0, 1, ..., L$:

$$\begin{aligned}
&\varepsilon_{-1}^{(l)} \equiv 0 \\
&\varepsilon_{0}^{(l)} \equiv S_l(\eta),\ l = 0, ..., L \\
&\varepsilon_{k+1}^{(l)} = \varepsilon_{k-1}^{(l+1)} + \left[ \varepsilon_{k}^{(l+1)} - \varepsilon_{k}^{(l)} \right]^{-1},\ k = 0, ..., L\ ;\ l = 0, ..., L - k - 1.
\end{aligned}$$

The recurrence forms a tableau

$$
\begin{array}{cccccc}
\varepsilon_0^{(0)} & \varepsilon_1^{(0)} & \varepsilon_2^{(0)} & \cdots & \varepsilon_{L-1}^{(0)} & \varepsilon_L^{(0)} \\
\varepsilon_0^{(1)} & \varepsilon_1^{(1)} & \varepsilon_2^{(1)} & \cdots & \varepsilon_{L-1}^{(1)} & \\
\varepsilon_0^{(2)} & \cdots & & \cdots & & \\
\cdots & & \varepsilon_2^{(L-2)} & & & \\
& \varepsilon_1^{(L-1)} & & & & \\
\varepsilon_0^{(L)} & & & & &
\end{array}
\tag{8}
$$

where each element of an even column estimates the limit. Convergence comes from interrogation of the last term of even columns $\varepsilon_i^{(L-i)},\ i = 0,2,...,2[L/2]$

$$
e_{We} \equiv \left| \frac{\varepsilon_i^{(L-i)} - \varepsilon_i^{(L-i-2)}}{\varepsilon_i^{(L-i)}} \right| < \varepsilon,\ i = 2,...,2[L/2]. \tag{9}
$$

Therefore, with $S_l$ appropriately defined, the limits of the sequences

$$
\begin{aligned}
f_l(\eta,\alpha) &= \sum_{k=0}^{l} a_k(0,\alpha)\eta^k \\
f_l'(\eta,\alpha) &= \sum_{k=0}^{l} k a_k(0,\alpha)\eta^{k-1} \\
f_l''(\eta,\alpha) &= \sum_{k=0}^{l} k(k-1) a_k(0,\alpha)\eta^{k-2}
\end{aligned}
\tag{10}
$$

give the desired solution, provided all converge, which also has not been shown, but is obviously assumed.

As a verification of convergence acceleration for $\beta \geq 0$, we first consider the Blasius solution [ $\beta = 0$ ] since there is a known highly accurate value of $\alpha$ for this case. In particular, the shooting angle from an extensive Runge-Kutta calculation is

$$
\alpha = 0.33205733621519630. \tag{11}
$$

to 17 digits [7]. Then, assuming this value initiate the recurrence relation of Eq(4a) with Eqs(4b) gives Fig. 1 for the convergence of the velocity profile partial sums $f_l'$ *without* (a) and *with* (b) *We* acceleration. The effectiveness of *We* acceleration is quite apparent. As $\eta$ approaches the radius of convergence, the original partial sums begins to oscillate while the *We* converged sequence remains steady and accurate. If we allow $\eta$ to become larger than radius of convergence, Fig. 2 results for $|1-f'(\eta)|$. The original sequence is now entirely out of numerical control. However, the *We* accelerated sequence converges to 5 or 6 places of the correct asymptotic value even when the original sums are take values of $10^{14}$ or $10^{15}$. This is a remarkable demonstration, in general, of the power of convergence acceleration.

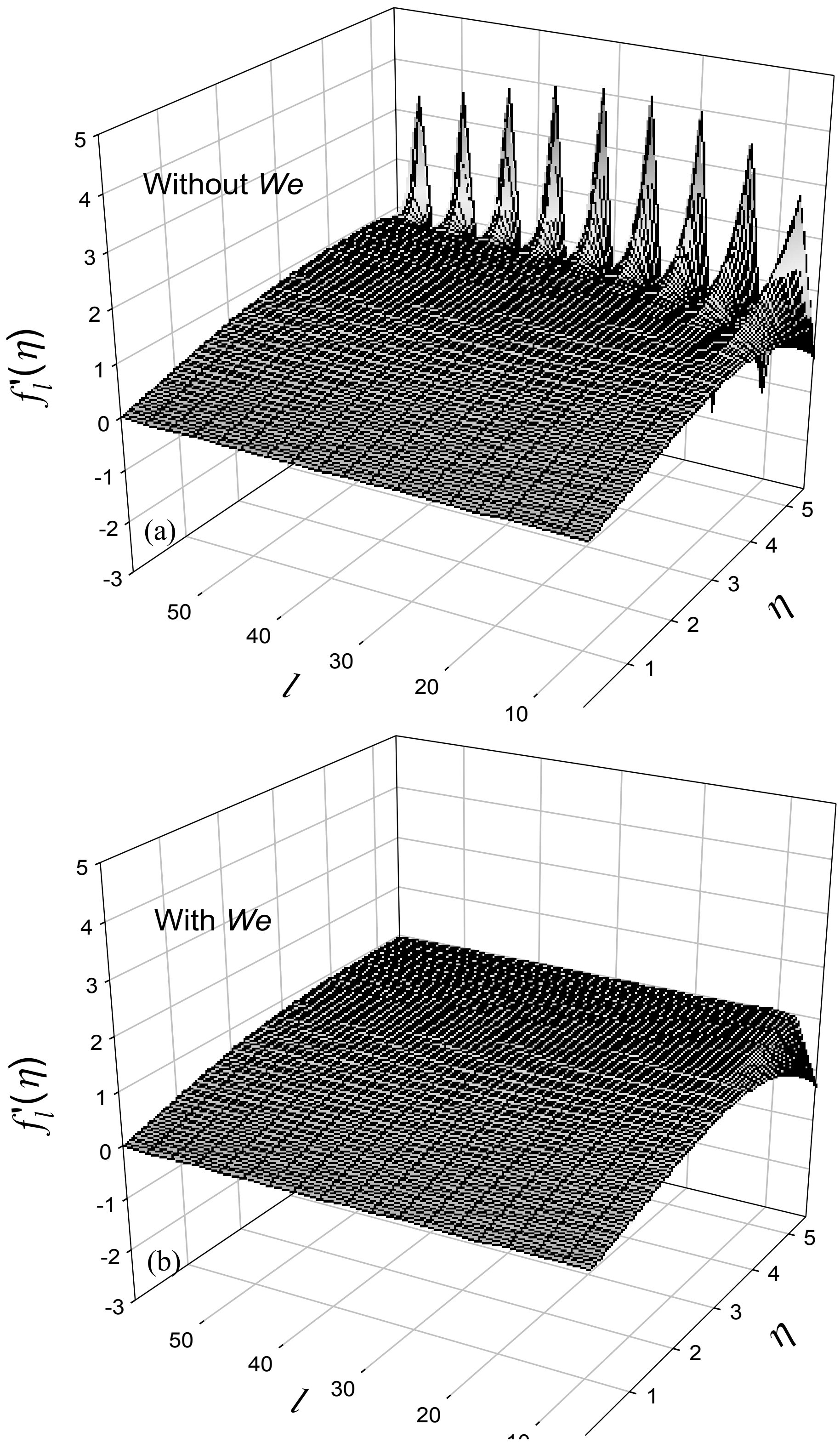

Fig. 1. Sequence convergence for the Blasius solution for the velocity profile from Eq(5) *without* (a) and *with* (b) *We* acceleration.

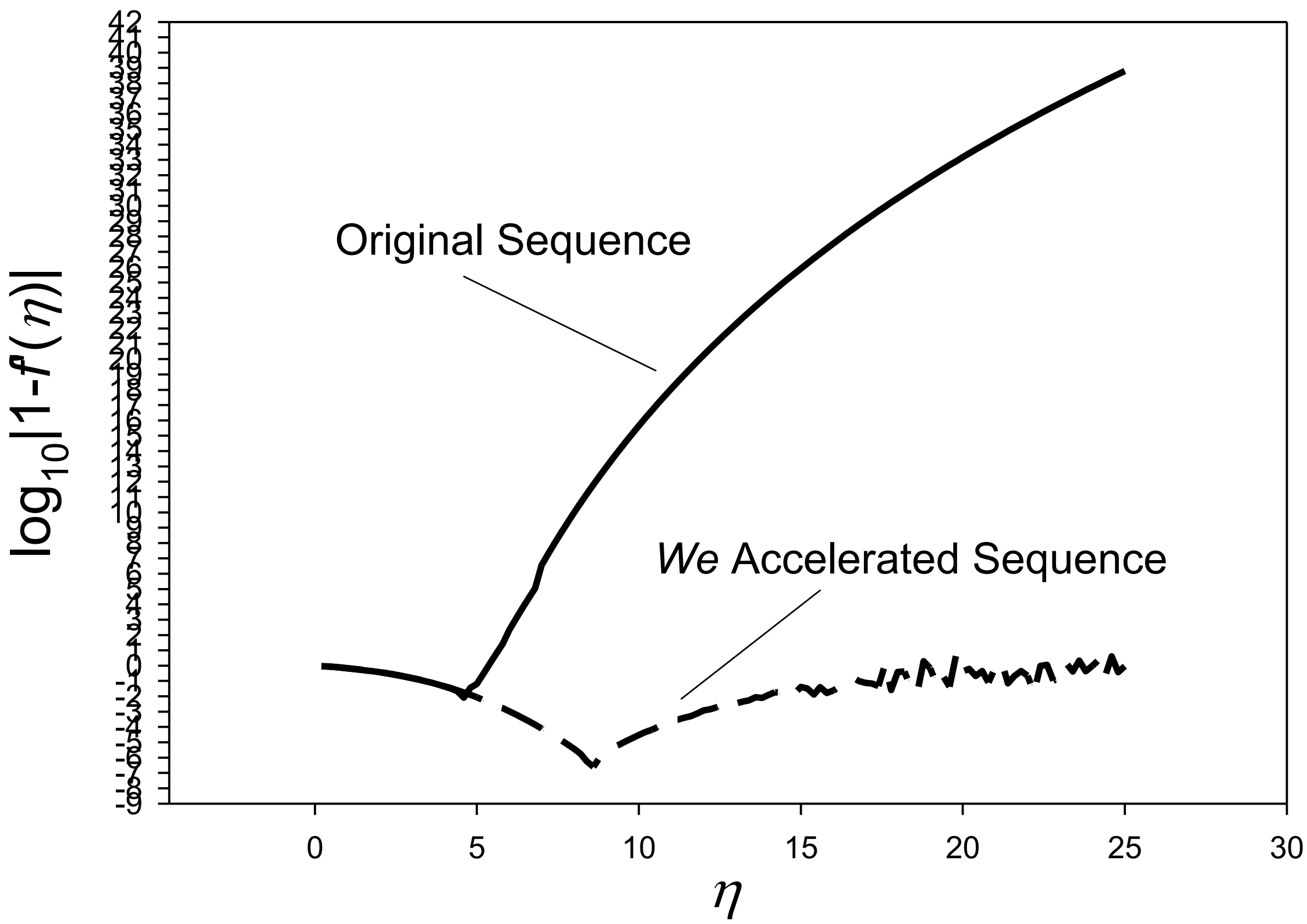

Fig. 2a. Velocity profile $f'(\eta,\alpha)$ from Eq(5).

However, eventually (for $\eta$ > 8), even the *We* sequence fails to converge. Therefore, to overcome this difficulty, we propose continuous analytical continuation.

### 3.2 Continuous analytical continuation (CAC)

CAC [6] is the continuation of a Taylor series into the complex plane, or in this case along the real axis. This is accomplished by re-starting the series at successive

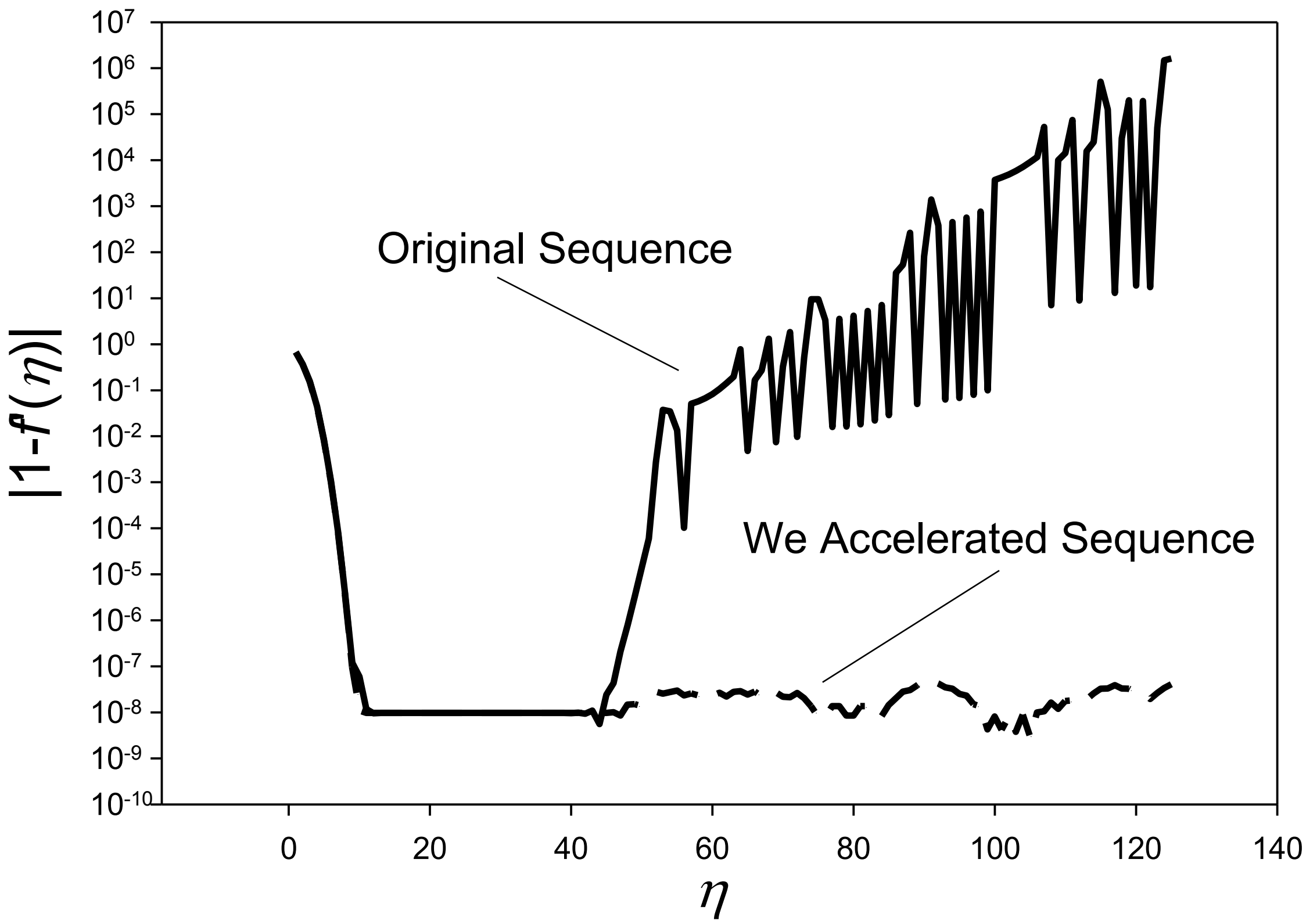

Fig, 2b. Velocity profile $f'(\eta,\alpha)$ from CAC.

intervals, $\Delta\eta$. Thus, Eq(4a) remains valid but now the initial three derivatives are

$$
\begin{aligned}
a_0(\eta_0,\alpha) &= f(\eta_0,\alpha) \\
a_1(\eta_0,\alpha) &= f'(\eta_0,\alpha) \\
a_2(\eta_0,\alpha) &= f''(\eta_0,\alpha)/2,
\end{aligned}
\qquad (4b')
$$

which replace Eq(4b). $\eta_0$ is the initial point of the interval, and $f(\eta_0,\alpha), f'(\eta_0,\alpha), f''(\eta_0,\alpha)$ are presumably available from the end of the previous interval. The desired quantities are therefore

$$f(\eta_0+\Delta\eta,\alpha)=\sum_{k=0}^{\infty} a_k(\eta_0,\alpha)(\Delta\eta)^k$$

$$f'\left(\eta_0+\Delta\eta,\alpha\right)=\sum_{k=0}^{\infty}ka_k\left(\eta_0,\alpha\right)\left(\Delta\eta\right)^{k-1}. \tag{12}$$

$$f''\left(\eta_0+\Delta\eta,\alpha\right)=\sum_{k=0}^{\infty}k\left(k-1\right)a_k\left(\eta_0,\alpha\right)\left(\Delta\eta\right)^{k-2}.$$

With CAC applied to the case of Fig 2a, Fig 2b results. Now the unaccelerated (original) partial sums converge up to $\eta$ = ~45, but eventually fail. The *We* converged sequence however, maintains the correct value to $\eta$ = 125 and beyond—again a remarkable demonstration of *We* acceleration. It would certainly be a challenge for a Runge-Kutta based method, or any other existing method for that matter, to reproduce this degree of accuracy. In particular, the challenge is for a numerical method, with no more than 16-digit precision and a mesh spacing of 1, to give the Blasius velocity profile at unity to eight digits of accuracy for $\eta$ greater than 100. The accuracy of the evaluation of Eq(5) at the initial point of the interval via converged acceleration, is the key to the success of the evaluation.

## 4. THE SHOOTING ALGORITHM

We now formulate the conventional shooting algorithm. The concept is to assume the shooting angle $\alpha$ and iterate until the condition

$$\lim_{\eta\to\infty}f'\left(\eta,\alpha\right)=1 \tag{13}$$

is satisfied to an acceptable accuracy. One of the difficulties of this approach is the sensitivity of the velocity profile to $\alpha$. We develop the algorithm for $\beta\geq 0$ and $\beta<0$ separately.

### 4.1 For $\beta\geq 0$

Consider Homann flow near the stagnation point of the boundary layer on a surface of revolution. For this case, $\beta_0=2,\ \beta=1$, where a 10-place value of $\alpha$ is known to be [14]

$$\alpha=1.3119376939. \tag{14}$$

Figure 3a shows the velocity profile as $\alpha$ varies from 2 to 0. As we pass through the correct $\alpha$ [Eq(14)], the profile takes on a value of unity at large $\eta$, which is more evident in the tight view in Fig 3b. This behavior seems typical of the general case for $\beta$ > 0 and is true for all the physically relevant (and physical) $\beta$ investigated. The only differences occur for $\eta$ usually under 10. In particular, we pass through the asymptotic velocity profile by decreasing $\alpha$ from above as has been shown numerically for positive $\beta$ [9]. In addition, the asymptotic profile is approached from below without overshoot [9]. While this has not been shown for all positive values of $\beta_0$, from this author's experience it seems true and therefore is assumed until circumstances dictate otherwise. The shooting algorithm therefore consists of the following procedure:

a. Starting from a relatively large value of $\alpha$ as the initial guess, $f'(\eta)$ is evaluated by increasing $\eta$ through steps of $h$ from zero to $\eta_m$.
b. If at some $\eta$, $f'(\eta)>1$, then $\alpha$ is decreased and $f'(\eta)$ evaluated until $f'(\eta)<1$ for some $\alpha$. At this point, the asymptotic profile is bracketed.
c. A new $\alpha$ is then determined by bisection and interrogation of the profile continues with subsequent $\alpha$'s found by bisection as $f'(\eta)$ oscillates about unity.
d. If $f'(\eta)$ does not cross unity from below as $\eta$ increases from zero to $\eta_m$, then $f''(\eta)$ is checked for negativity. If negative, $\alpha$ is below its correct value and bisection again determines the next $\alpha$. Experience indicates that $\eta_m = 20$ is sufficient for $\beta > 0$.
e. Finally, when the estimate for $\alpha$ is approximately within an order of magnitude of the desired error, we apply the secant method rather than bisection.

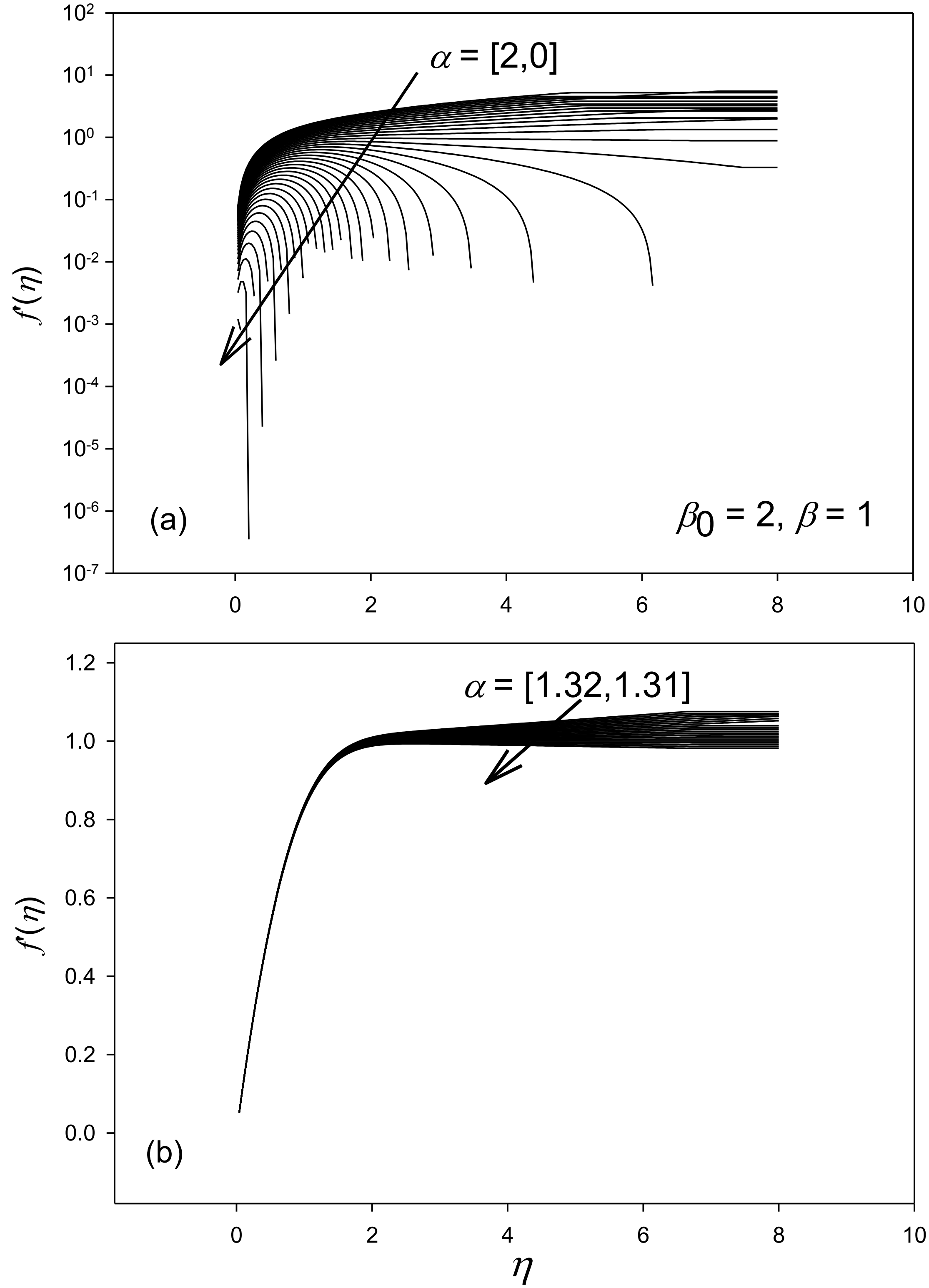

Fig 3. Variation of velocity profile with $\alpha$ for Homann flow:
(a) Wide view (b) Tight view.

Table 1 gives results for the above procedure applied to Homann flow. The entry for error of $10^{-12}$ confirms the value quoted by Zhang and Chen [14] over that of Asaithambi [6]. Also indicated is the number of (bisection) iterations for convergence. It should be noted that bisection is the preferred method over Newton-Raphson to find the correct $\alpha$ because of its reliability. Hence, the security of bisection is more highly valued than the faster convergence of Newton-Raphson. In this way, potential instability is avoided. In any case, to obtain the accuracy of the recent literature, less than 30 iterations are required. By all accounts, about 50 iterations are typical for 14-digit accuracy.

As a further indication of the power of the convergence acceleration, a case, at an extreme accuracy of $10^{-32}$ (in Quad Precision arithmetic) is included. 29 digits are expected to be accurate (last entry) as verified by the Pohlhausen flow example considered next.

**Table 1**
Variation of a for Homann Flow with the specified error

| $\varepsilon$ | $\alpha$ | *#Itr* |
|---|---|---|
| $10^{-8}$ | 1.3119377 | 24 |
| $10^{-9}$ | 1.31193769 | 28 |
| $10^{-10}$ | 1.311937690 | 33 |
| $10^{-11}$ | 1.3119376938 | 34 |
| $10^{-12}$ | 1.31193769392 | 40 |
| $10^{-13}$ | 1.31193769388 | 39 |
| $10^{-14}$ | 1.3119376938798 | 44 |
| $10^{-32}$ | 1.3119376938798051354816461707 | 104(QP) |

Pohlhausen flow, for which $\beta_0 = 0,\ \beta = 1$, is a rare instance of an analytical solution to the F-S profiles and for $\alpha$, which is $2/\sqrt{3}$. As shown in Table 2, the proposed algorithm gives 15 correct places for a requested error of $10^{-16}$ and 29 (QP) for an error of $10^{-32}$. Arguably, this example provides a most convincing demonstration of convergence acceleration coupled to continuous analytical continuation.

**Table 2**
Pohlhausen flow $\beta_0 = 0,\ \beta = 1$

| $\varepsilon$ | $\alpha$ |
|---|---|
| $10^{-16}$ | 1.154700538379252 |
| $10^{-32}$ | 1.15470053837925152901829756101 |
| Exact | 1.15470053837925152901829756100 |

Before addressing the standard cases found in the literature, we consider the Blasius solution in some detail. Table 3a gives highly accurate values of $\alpha$ and Table 3b gives the Blasius function and its derivatives to 10-digits. These values are in exact agreement with the 8-place results of in Ref. 15, which seems to be the most accurate published to date. Please note that the values in this reference were obtained from an integral formulation with Eq(5) at its core in 1981. In addition,

**Table 3a**

Blasius Flow $\beta_0 = 1,\ \beta = 0$

| | $\alpha$ | **#*Itr*** |
|---|---|---|
| *Exact* | 0.33205733621519630 | |
| $\varepsilon = 10^{-16}$ | 0.332057336215195 | 52 |
| $\varepsilon = 10^{-32}$ | 0.33205733621519629893718006201 | 104 (QP) |

the asymptotic value

$$\lim_{\eta\to\infty}\left[\eta - f(\eta)\right]$$

is 1.7207876575205 in DP arithmetic agreeing to 14-digits with the value of Ref. 3. In QP arithmetic, we find 1.720787657520502812, which agrees to all digits when rounded and adds several. The degree of agreement for a calculation without multiple precision arithmetic beyond QP is rather notable and should be the standard to which algorithms are held.

All calculations for the above 3 figures and 3 tables required under 3 minutes on a Gateway 1.4 GHz laptop.

### 4.2 For $\beta_{\min} \le \beta < 0$

Figure 4 gives the variation of the velocity profile as $\alpha$ varies from 1 to 0 for $\beta = -0.1$. As for positive $\beta$, when $\alpha$ passes through its correct value [9], the velocity takes on an its asymptotic profile. However, unlike for positive $\beta$, the velocity profile initially overshoots its asymptotic limit. Even with this difference, the shooting strategy outlined above remains valid since the approach is still from above.

## 5. COMPARISONS OF RESULTS

In this section, we compare the results of the present algorithm for the F-S solution with some of the most accurate results found in the literature. The physical flows, forward and reverse, are considered separately.

### 5.1 Forward flows

Table 4a lists converged $\alpha$ for positive $\beta$ to 40. When rounded and compared to the corresponding results of Refs. 14, 16 and 17, we find all values in agreement to 6-places except for the last digit as indicated in Table 4b. In addition, the free boundary $\eta_\infty$ layer estimate, defined here when the velocity profile becomes unity

**Table 3b**
Blasius Flow Profiles

| $\eta$ | $f$ | $f'$ | $f''$ |
|---|---|---|---|
| 0.0E+00 | 0.000000000E+00 | 0.000000000E+00 | 3.320573362E-01 |
| 2.0E-01 | 6.640999715E-03 | 6.640779210E-02 | 3.319838371E-01 |
| 4.0E-01 | 2.655988402E-02 | 1.327641608E-01 | 3.314698442E-01 |
| 6.0E-01 | 5.973463750E-02 | 1.989372524E-01 | 3.300791276E-01 |
| 8.0E-01 | 1.061082208E-01 | 2.647091387E-01 | 3.273892701E-01 |
| 1.0E+00 | 1.655717258E-01 | 3.297800312E-01 | 3.230071167E-01 |
| 1.2E+00 | 2.379487173E-01 | 3.937761044E-01 | 3.165891911E-01 |
| 1.4E+00 | 3.229815738E-01 | 4.562617647E-01 | 3.078653918E-01 |
| 1.6E+00 | 4.203207655E-01 | 5.167567844E-01 | 2.966634615E-01 |
| 1.8E+00 | 5.295180377E-01 | 5.747581439E-01 | 2.829310173E-01 |
| 2.0E+00 | 6.500243699E-01 | 6.297657365E-01 | 2.667515457E-01 |
| 2.2E+00 | 7.811933370E-01 | 6.813103772E-01 | 2.483509132E-01 |
| 2.4E+00 | 9.222901256E-01 | 7.289819351E-01 | 2.280917607E-01 |
| 2.6E+00 | 1.072505977E+00 | 7.724550211E-01 | 2.064546268E-01 |
| 2.8E+00 | 1.230977302E+00 | 8.115096232E-01 | 1.840065939E-01 |
| 3.0E+00 | 1.396808231E+00 | 8.460444437E-01 | 1.613603195E-01 |
| 3.2E+00 | 1.569094960E+00 | 8.760814552E-01 | 1.391280556E-01 |
| 3.4E+00 | 1.746950094E+00 | 9.017612214E-01 | 1.178762461E-01 |
| 3.6E+00 | 1.929525170E+00 | 9.233296659E-01 | 9.808627878E-02 |
| 3.8E+00 | 2.116029817E+00 | 9.411179967E-01 | 8.012591814E-02 |
| 4.0E+00 | 2.305746418E+00 | 9.555182298E-01 | 6.423412109E-02 |
| 4.2E+00 | 2.498039663E+00 | 9.669570738E-01 | 5.051974749E-02 |
| 4.4E+00 | 2.692360938E+00 | 9.758708321E-01 | 3.897261085E-02 |
| 4.6E+00 | 2.888247990E+00 | 9.826835008E-01 | 2.948377201E-02 |
| 4.8E+00 | 3.085320655E+00 | 9.877895262E-01 | 2.187118635E-02 |
| 5.0E+00 | 3.283273665E+00 | 9.915419002E-01 | 1.590679869E-02 |
| 5.2E+00 | 3.481867612E+00 | 9.942455354E-01 | 1.134178897E-02 |
| 5.4E+00 | 3.680919063E+00 | 9.961553040E-01 | 7.927659815E-03 |
| 5.6E+00 | 3.880290678E+00 | 9.974777682E-01 | 5.431957680E-03 |
| 5.8E+00 | 4.079881939E+00 | 9.983754937E-01 | 3.648413667E-03 |
| 6.0E+00 | 4.279620923E+00 | 9.989728724E-01 | 2.402039844E-03 |
| 6.2E+00 | 4.479457297E+00 | 9.993625417E-01 | 1.550170691E-03 |
| 6.4E+00 | 4.679356615E+00 | 9.996117017E-01 | 9.806151170E-04 |
| 6.6E+00 | 4.879295811E+00 | 9.997678702E-01 | 6.080442648E-04 |
| 6.8E+00 | 5.079259772E+00 | 9.998638190E-01 | 3.695625701E-04 |
| 7.0E+00 | 5.279238811E+00 | 9.999216041E-01 | 2.201689553E-04 |
| 7.2E+00 | 5.479226847E+00 | 9.999557173E-01 | 1.285698072E-04 |
| 7.4E+00 | 5.679220147E+00 | 9.999754577E-01 | 7.359298339E-05 |
| 7.6E+00 | 5.879216466E+00 | 9.999866551E-01 | 4.129031111E-05 |
| 7.8E+00 | 6.079214481E+00 | 9.999928812E-01 | 2.270775140E-05 |
| 8.0E+00 | 6.279213431E+00 | 9.999962745E-01 | 1.224092624E-05 |
| 8.2E+00 | 6.479212887E+00 | 9.999980875E-01 | 6.467978611E-06 |
| 8.4E+00 | 6.679212609E+00 | 9.999990369E-01 | 3.349939753E-06 |
| 8.6E+00 | 6.879212471E+00 | 9.999995242E-01 | 1.700667989E-06 |
| 8.8E+00 | 7.079212403E+00 | 9.999997695E-01 | 8.462841214E-07 |

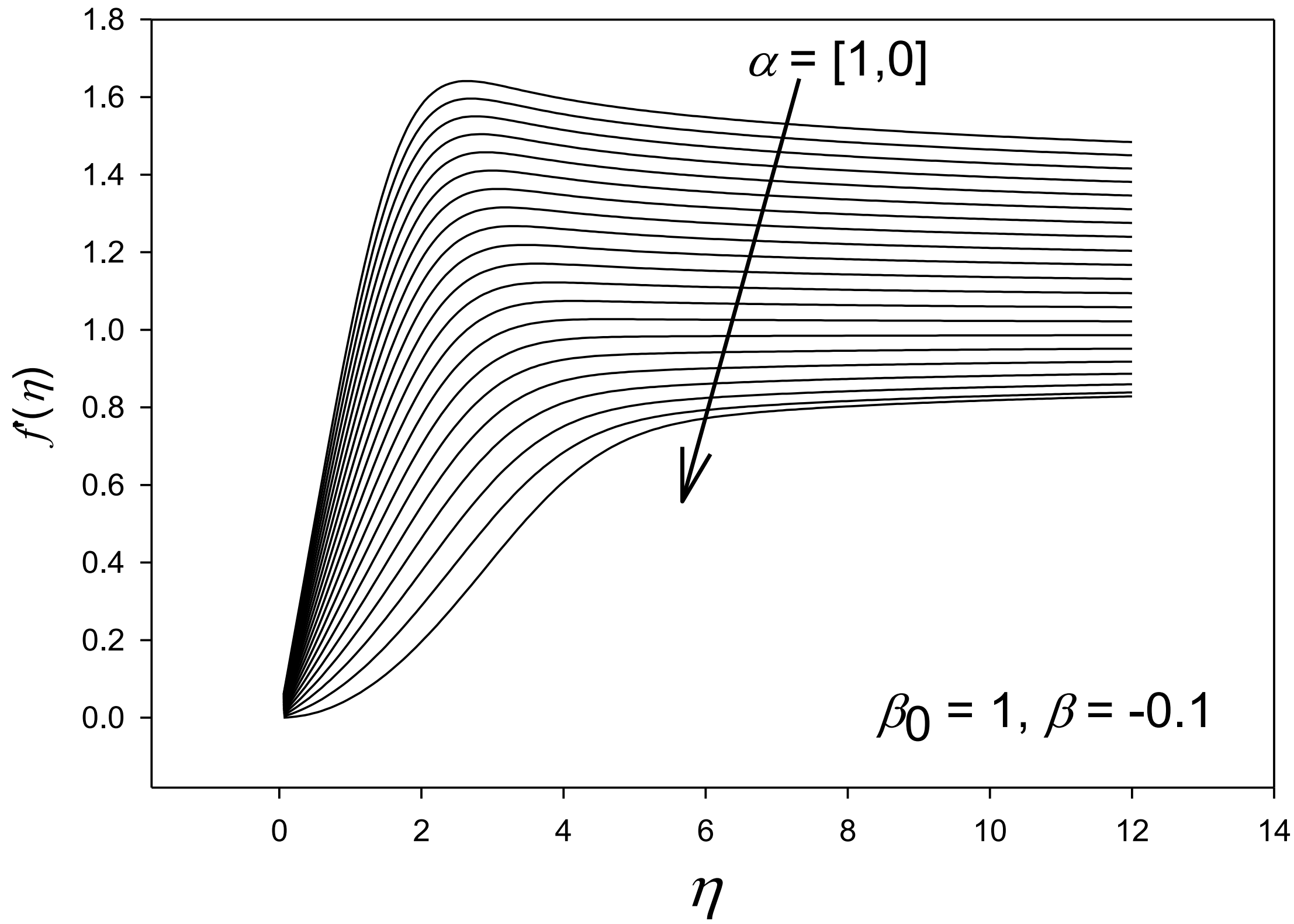

Fig.4. Variation of the velocity profile for $\beta$ = -0.1.

**Table 4a**

For $0 \leq \beta$

| $\boldsymbol{\beta}$ | $\boldsymbol{\alpha}$ | $\eta_\infty$ | **#*Itr*** |
|---|---|---|---|
| 40 | 7.31478497433 | 1.57-1.58 | 49 |
| 30 | 6.33820862834 | 1.79-1.80 | 57 |
| 20 | 5.18071802491 | 2.14-2.15 | 50 |
| 15 | 4.49148689764 | 2.42-2.43 | 48 |
| 10 | 3.67523410111 | 2.83-2.84 | 53 |
| 2 | 1.68721816921 | 4.46-4.47 | 49 |
| 1 | 1.23258765682 | 4.98-4.99 | 53 |
| 0.5 | 0.927680039837 | 5.37-5.38 | 50 |
| 0 | 0.469599988361 | 6.07-6.08 | 52 |

**Table 4b**
Last digits not in agreement in Refs. 14,16 and 17 for $\beta$s indicated

| | Ref | $\beta$ |
|---|---|---|
| $\beta \geq 0$ | 14 | 1, 15, 30 |
| | 16 | 1 |
| $\beta < 0$ | 17 | -0.15 |
| | 16 | -0.18 |
| | 17 | -0.15, -0.18 |

**Table 5**
Forward flow for $-0.198837735 \leq \beta < 0$

| $\beta$ | $\alpha$ | $\eta_\infty$ | *Itr* |
|---|---|---|---|
| -0.10 | 3.19269759843 | 6.36-6.37 | 48 |
| -0.15 | 2.16361405647 | 6.60-6.61 | 48 |
| -0.18 | 1.28636220596 | 6.85-6.86 | 49 |
| -0.1988 | 5.218187884E-03 | 7.32-7.33 | 42 |
| -0.198837 | 7.24675233 E-04 | 7.32-7.33 | 59 |
| -0.1988377 | 1.58136616 E-04 | 7.35-7.36 | 91QP |
| -0.198837735 | 5.77016 E-06 | 7.35-7.36 | 95QP |

to 5 places, is also given in Table 4a. Finally, as a measure of the computational effort, the number of bisections for DP arithmetic is also provided.

Table 5 gives results for $\beta_{\min} \leq \beta < 0$. For the first three entries, all values agree to the fifth place with Refs. 14, 16 and 17 except as again indicated in Table 4b. For the fourth entry, only the first 2 digits agree however. This value of $\beta$ is near the boundary between existence and non-existence ($\beta_{\min}$) of the solution and is particularly difficult numerically troubling. The value quoted in the table is expected to be accurate to all but the last digit. The next three entries are an attempt to more accurately define this boundary. For these $\beta$'s, the determination of $\alpha$ requires QP arithmetic and further independent verification.

The final comparison for forward flows is with a relatively obscure but very important data set. Apparently, M. Katagiri [18], in a symposium on space flight held at the University of Tokyo, published what must be considered the most accurate determination of $\alpha$ before the present investigation. Using a quasi-linear approximation in an integral formulation, Katagiri was able to compute values of $\alpha$ for $-0.19 \le \beta \le 1$ to 10 or 11 digits. Unfortunately, this work has been lost in antiquity, but its accuracy has been resurrected here in Table 6 for $\alpha$ with two additional digits as found from the present algorithm. All 40 values are in roundoff agreement with Katagiri's results. The table required 2 min of computational time on a Gateway 1.4 GHz laptop. This agreement makes the work of Katagiri that much more impressive as Katagiri published his results in 1986-- during the era of primitive computing. In addition, had this reference been more widely known in the West, many of the recent numerical approaches would never have been attempted, as they are far inferior. It is only fitting to acknowledge Katagiri as the first numerical analyst to achieve truly high precision results for the solution of the F-S equation.

### 5.1 Reverse flows

A second solution of the F-S equation exists in the range $\beta_{\min} \le \beta < 0$ representing reverse flow. Figure 5 shows the variation of the velocity profile now for a negative shooting angle for $\beta = -0.01$ and $-0.1$. As above, the profile passes through the asymptotic condition of unity, from below with increasing $\alpha$. This behavior is again captured by a bisection procedure once the asymptotic limit is first surpassed. Table 7 gives $\alpha$- values for reversed flow for the $\beta$-values considered by Katagiri in Table 6. This table required 110s of computing time and should be correct to several units in the last place.

In the interest of adding an ending to a previous investigation on reverse flow, we consider one of the earliest and most influential articles regarding the shooting method [9]. In Table 8, the Cebeci and Keller investigation is completed. In Ref. 9, only 3 digits for the given $\beta$'s were found for reverse flow. Except for the last 3 entries, the 3 digits are in nearly complete agreement with the present calculation. The last entry was modified since the value in Ref. 9 was in the region of no solution. It should be noted that the value close to the existence boundary on the lower branch nearly matches that of the upper branch as expected. Figure 6 displays the velocity profiles for the cases of Ref. 9. The region of reversed flow is clearly evident.

## 6. CONCLUSION

An efficient algorithm, arguably the most efficient to date, has been devised to resolve the Falkner-Skan equation of boundary layer theory. The algorithm is simple in that it is based on a Maclaurin series solution through recurrence. Most

**Table 6**

Comparison to Katagiri Results

| $\beta$ | $\alpha$ |
|---|---|
| 1.00E+00 | 1.23258765682E+00 |
| 9.50E-01 | 1.20546125458E+00 |
| 9.00E-01 | 1.17772781917E+00 |
| 8.50E-01 | 1.14934554396E+00 |
| 8.00E-01 | 1.12026765738E+00 |
| 7.50E-01 | 1.09044156217E+00 |
| 7.00E-01 | 1.05980777320E+00 |
| 6.50E-01 | 1.02829859292E+00 |
| 6.00E-01 | 9.95836440616E-01 |
| 5.50E-01 | 9.62331717602E-01 |
| 5.00E-01 | 9.27680039837E-01 |
| 4.50E-01 | 8.91758591637E-01 |
| 4.00E-01 | 8.54421231190E-01 |
| 3.50E-01 | 8.15491778664E-01 |
| 3.00E-01 | 7.74754580311E-01 |
| 2.50E-01 | 7.31940848513E-01 |
| 2.00E-01 | 6.86708181032E-01 |
| 1.50E-01 | 6.38608512560E-01 |
| 1.00E-01 | 5.87035219198E-01 |
| 5.00E-02 | 5.31129630465E-01 |
| 0.00E+00 | 4.69599988361E-01 |
| -1.00E-02 | 4.56454824718E-01 |
| -2.00E-02 | 4.42982276168E-01 |
| -3.00E-02 | 4.29156500469E-01 |
| -4.00E-02 | 4.14947955734E-01 |
| -5.00E-02 | 4.00322595418E-01 |
| -6.00E-02 | 3.85240819783E-01 |
| -7.00E-02 | 3.69656086478E-01 |
| -8.00E-02 | 3.53513033028E-01 |
| -9.00E-02 | 3.36744882009E-01 |
| -1.00E-01 | 3.19269759843E-01 |
| -1.10E-01 | 3.00985311136E-01 |
| -1.20E-01 | 2.81760524240E-01 |
| -1.30E-01 | 2.61422755987E-01 |
| -1.40E-01 | 2.39735955312E-01 |
| -1.50E-01 | 2.16361405647E-01 |
| -1.60E-01 | 1.90779855269E-01 |
| -1.70E-01 | 1.62114677053E-01 |
| -1.80E-01 | 1.28636220596E-01 |
| -1.90E-01 | 8.56997440597E-02 |

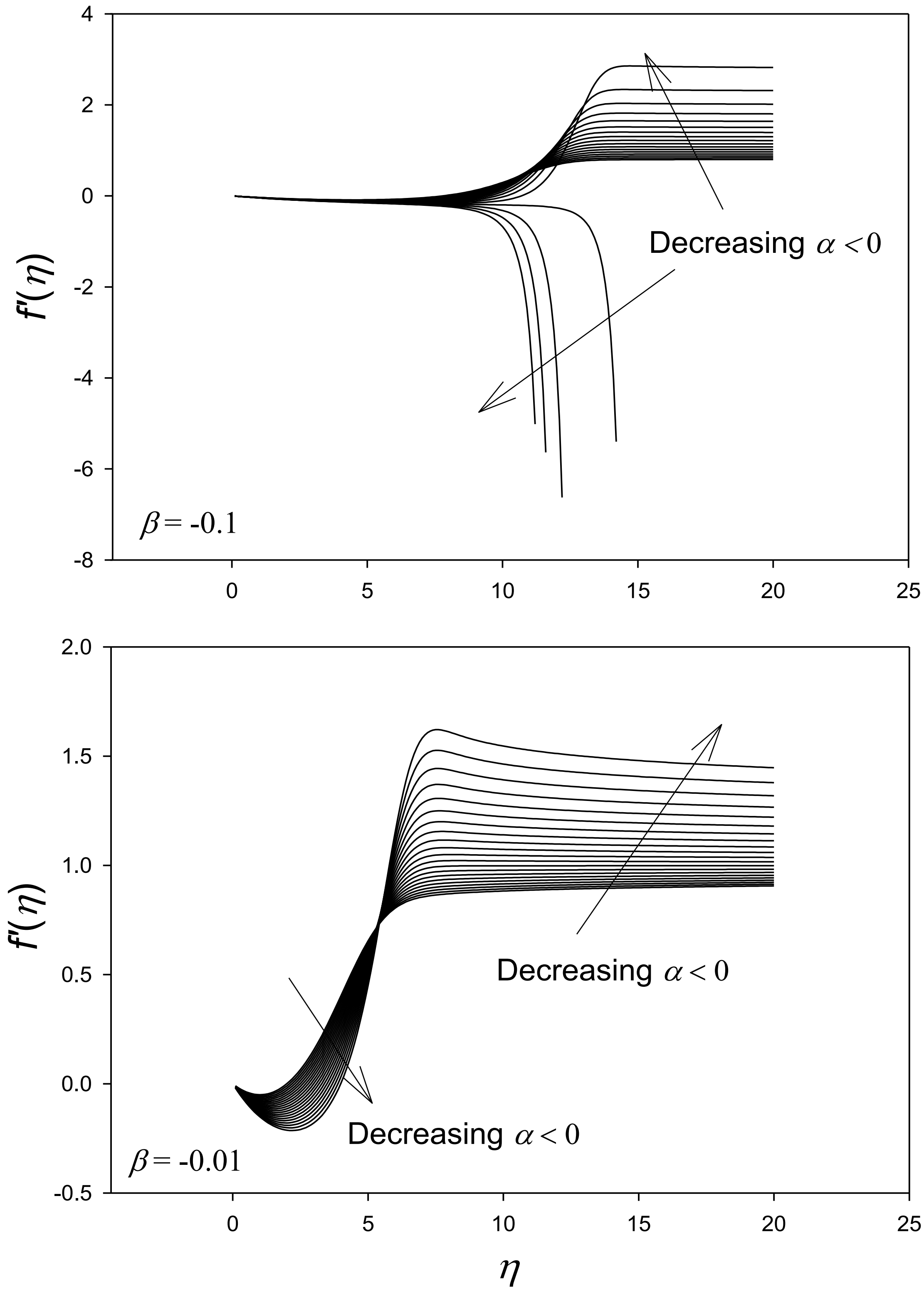

Fig. 5. Variation of the velocity profile for reversed flow for $\beta = -0.01$ and $-0.1$.

**Table 7**
Reversed flow for $0.198837735 < \beta < 0$

| $\beta$ | $\alpha$ |
|---|---|
| -1.00E-02 | -4.23209260392E-02 |
| -2.00E-02 | -6.51685855429E-02 |
| -3.00E-02 | -8.25629651188E-02 |
| -4.00E-02 | -9.66367874086E-02 |
| -5.00E-02 | -1.08271083286E-01 |
| -6.00E-02 | -1.17924111038E-01 |
| -7.00E-02 | -1.25858531722E-01 |
| -8.00E-02 | -1.32227654380E-01 |
| -9.00E-02 | -1.37113811802E-01 |
| -1.00E-01 | -1.40546212979E-01 |
| -1.10E-01 | -1.42507910309E-01 |
| -1.20E-01 | -1.42935194358E-01 |
| -1.30E-01 | -1.41709502152E-01 |
| -1.40E-01 | -1.38638925254E-01 |
| -1.50E-01 | -1.33421237895E-01 |
| -1.60E-01 | -1.25567664217E-01 |
| -1.70E-01 | -1.14227239098E-01 |
| -1.80E-01 | -9.76920601346E-02 |
| -1.90E-01 | -7.13359060034E-02 |

importantly, the a highly accurate series evaluation is enabled through a Wynn-epsilon convergence acceleration coupled to continuous analytical continuation. The coupling has been demonstrated to provide the solution far into the asymptotic region. We then follow a shooting strategy based on bisection to force the shooting angle $\alpha$ (skin friction coefficient) toward its correct value as the velocity relaxes to its asymptotic profile. There is no Newton-Raphson search to adjust and control; and as long as the shooting angle is above its expected value, the initial guess is arbitrary. For an example of robustness, consider $\beta$ = 1000. An initial value for the iteration of $\alpha$ = 200 gives a converged value of 36.5171968. The accuracy of this value however awaits the development of second algorithm of greater of equal accuracy--which will be a challenge if method developers continues to use differential equation solvers. The present algorithm also accommodates reversed flow an example of which is shown in Fig. 7 for $\beta$= -0.16, where the F-S functions are contrasted for forward and reversed flow.

The significance of the investigation presented here is that it closes the chapter on the development of highly accurate algorithms for the F-S equation. The next challenge is to develop the algorithm for the non-adiabatic case, including the energy equation. Also under consideration is the case for stretching surfaces with blowing and suction. Of particular note is that the Maclaurin series solution, as suggested by Blasius, has been vindicated as the solution representation of choice and may prove to be a viable approach for resolution of more comprehensive flow scenarios in the future.

**Table 8**
Completing calculations for Cebeci and Keller [9] investigation

| $\beta$ | $\alpha$ |
|---|---|
| -9.16200E-03 | -3.99994660387E-02 |
| -2.47890E-02 | -7.39993700921E-02 |
| -4.02860E-02 | -9.70005720161E-02 |
| -4.97450E-02 | -1.08000456249E-01 |
| -1.01763E-01 | -1.41000029373E-01 |
| -7.95960E-02 | -1.31999450790E-01 |
| -1.52118E-01 | -1.31999333488E-01 |
| -1.80553E-01 | -9.65623555697E-02 |
| -1.80552E-01 | -9.65644238587E-02 |
| -1.96348E-01 | -4.00005870180E-02 |
| -1.98826E-01 | -2.88367895808E-03 |
| -1.98837735E-01 | -5.77014 E-06 |

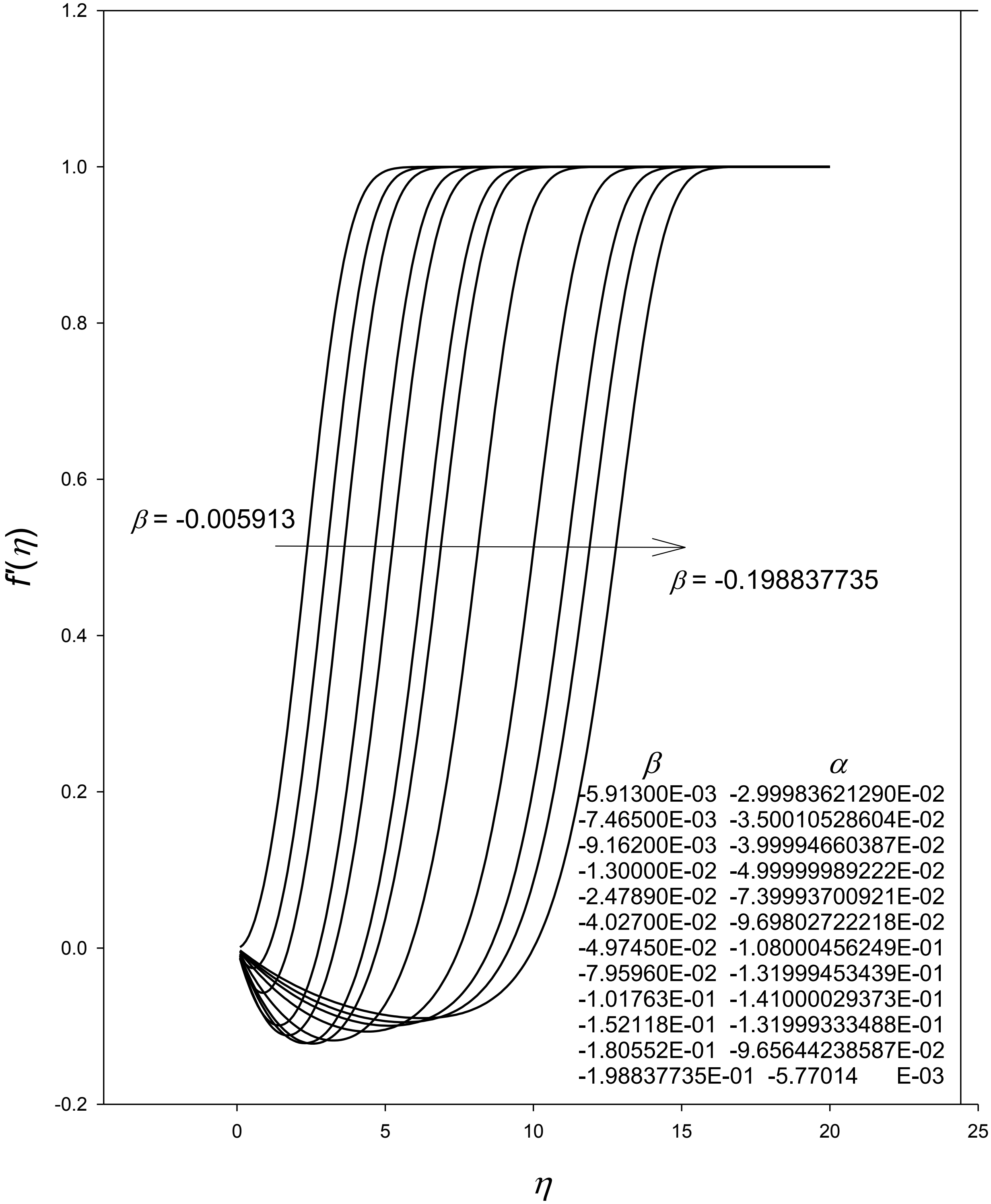

Fig. 6. Examples of reversed flow of Ref. 9.

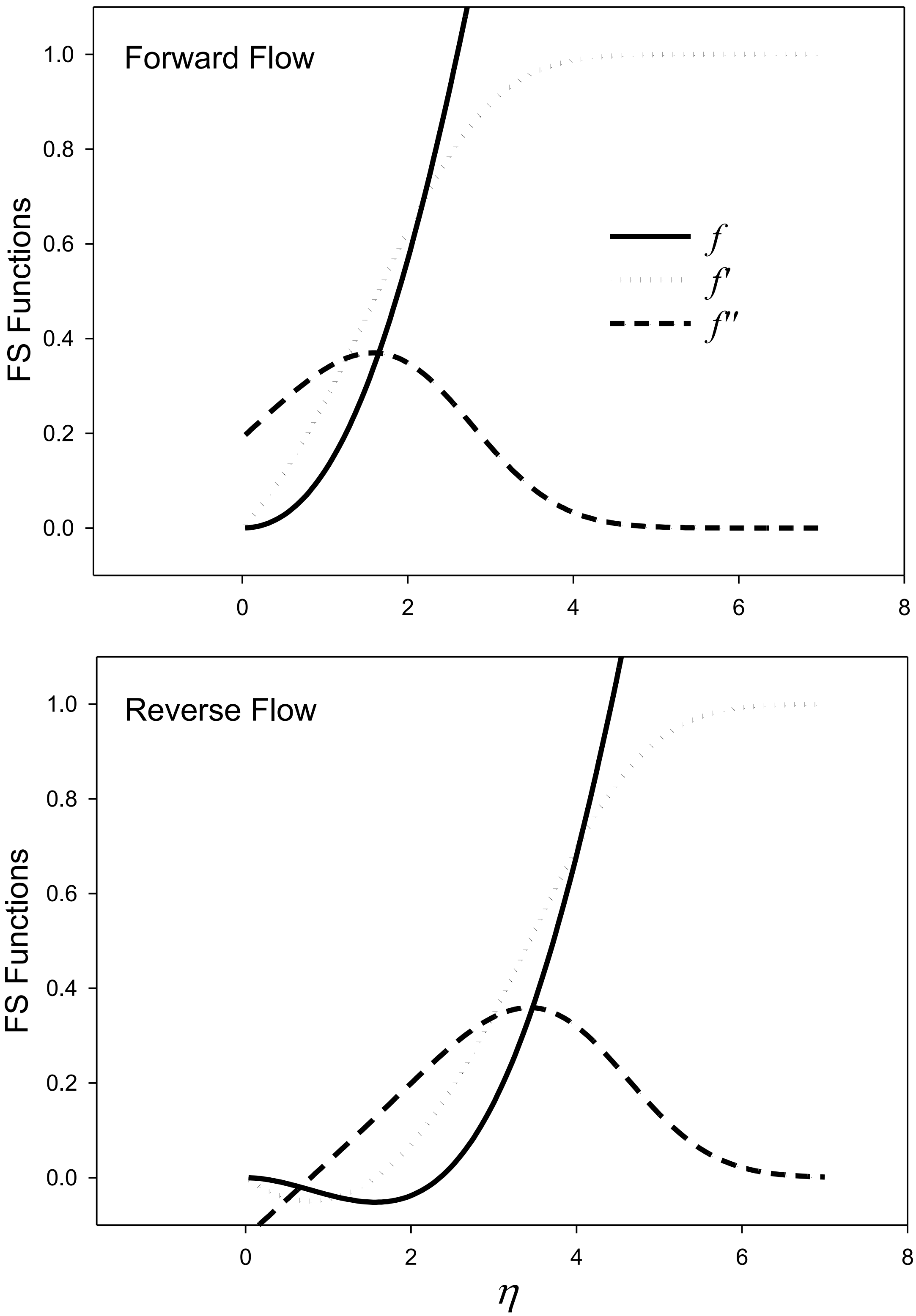

Fig. 7. Forward and reverse flow for $\beta$ = -0.16

**REFERENCES**

[1] H. Blasius, Grenrschichten in Flussigkeiten mit kleiner Reibung. *Z. Math, u. Phys.*, **56**, 1-37(1908).

[2] V.M. Falkner, S.W. Skan, Some approximate solutions of the boundary layer equations, *Philos. Mag.* **12**, 865–896(1931).

[3] J.P. Boyd, The Blasius Function : Computation Before Computers, the Value of Tricks, Undergraduate Projects, and Open Research Problems, *SIAM Review*, **50**, 791-804(2008).

[4] K. Parand a, M. Shahini a, Mehdi Dehghan, Solution of a laminar boundary layer flow via a numerical method, *Comm.. Nonlinear. Sci. Num. Sim.* 15 360–367(2010).

[5] URL Link: *Steven Finch November 12, 2008 The boundary value problem y000 (x)+ …*

[6] A. Asaithambi, Solution of the Falkner-Skan equation by recursive evaluation of Taylor coefficients, *J. Comput. Appl. Math*. **176** 203-214(2005).

[7] J. P. Boyd, The Blasius function in the complex plane, *Experiment. Math*. **8**, 381-394(1999).

[8] J. P. Boyd, Padé approximant algorithm for solving nonlinear ordinary differential equation boundary value problems on an unbounded domain, *Computers in Physics*, **11**, #3, 299-303(1997).

[9] T. Cebeci and H. B. Keller, Shooting and parallel shooting methods for solving the Falkner-Skan boundary-layer equation, *J. Comput. Phys*. **7**, 289-300(1971).

[10] P. Wynn, On a device for computing the $e_m(S_n)$ transformation, MATC, **10**, 91-96(1956).

[11] P. Wynn, On the Convergence and Stability of the Epsilon Algorithm, SIAM *J. Num, Anal*., **3**,#1 91-122(1966).

[12] N. Riley and P. D. Weidman, Multiple solutions of the Falkner-Skan equation for flow past a stretching boundary, *SIAM J. Appl. Math*. **49**, 1350-1358(1989).

[13] G. Baker and P. Graves-Morris (1996), *Pade' Approximants* (Cambridge University Press, NY).

[14] J. Zhang and B.Chen, An iterative method for solving the Falkner–Skan equation, *Applied Mathematics and Computation*, **210**, #1, 215-222(2009).

[15] J. Y. Parlange, R. D. Braddock, and G. Sander, Analytical Approximations to the Solution of the Blasius Equation, Acts Mechanics **38**, 119,125(1981).

[16] A. Asaithambi, Solution of the Falkner–Skan equation by recursive evaluation of Taylor coefficients, *J. Comput. Appl. Math.* **176**, 203–214(2005).

[17] A.A. Salama, Higher-order method for solving free boundary-value problems, *Numer. Heat Transfer, Part B: Fund.* **45**, 385–394(2004).

[18] M. Katagiri, On accurate numerical solutions of Falkner-Skan equation, *Proceedings of the Symposium on Mechanics for Space Flight*, Tokyo Univ, 1985.